\documentclass[preprint,3p,times]{elsarticle}

\usepackage{graphicx}
\usepackage{color}

\usepackage{amssymb}
\usepackage{amsmath}
\usepackage{amsthm}
\usepackage{lineno}
\usepackage{caption}
\usepackage{subcaption}

\usepackage{tikz}
\usetikzlibrary{shapes,arrows}

    \usepackage{framed} % Framing content
    \usepackage{multicol} % Multiple columns environment

    \usepackage{nomencl} % Nomenclature package
    
    \usepackage{ifthen}
    
    \makenomenclature
    \renewcommand*\nompreamble{\begin{multicols}{2}}
    \renewcommand*\nompostamble{\end{multicols}}
    
    \nomenclature[]{$W_0$}{melting velocity at the heat source center}
    \nomenclature[]{$W$}{melting velocity profile}
    \nomenclature[]{$p$}{pressure field}
    \nomenclature[]{$p_\text{eff}$}{exerted pressure}
    \nomenclature[]{$(i,j)$}{spatial node index}
    \nomenclature[]{$c_p$}{heat capacity}
    \nomenclature[]{$T$}{temperature field}
    \nomenclature[]{$A$}{area of the working surface}
    \nomenclature[]{$F_\text{eff}$}{exerted force}
    \nomenclature[]{$\hat{F}_\text{eff}$}{exerted force per unit length}
    \nomenclature[]{$F_p$}{pressure force}
    \nomenclature[]{$s$}{heat flux ratio}
    \nomenclature[]{$Q$}{heat flow rate}
    \nomenclature[]{$(P_1,P_2,P_3)$}{parameters for equation \eqref{eq:approximationVelocityDifference}}
    \nomenclature[G]{$\sigma$}{relaxation factor}
    \nomenclature[]{$q$}{heat flux}
    \nomenclature[]{$\text{Ste}$}{Stefan number}
    \nomenclature[]{$\text{Re}$}{Reynolds number}
    \nomenclature[]{$\text{Pe}$}{Peclet number}
    \nomenclature[]{$r_c$}{curve radius}
    \nomenclature[]{$(n_r,n_z)$}{number of nodes in (r,z) directions}
    \nomenclature[]{$N$}{total number of nodes}
    \nomenclature[]{$n$}{n=1 (\caseA), n=0 (\caseB)}
    \nomenclature[]{$R$}{Radius (\caseA) or half length (\caseB) of the heat source}
    \nomenclature[]{$h_m$}{latent heat of melting}
    \nomenclature[]{$h_m^*$}{reduced latent heat of melting}
    \nomenclature[]{$(r,z)$}{coordinates, see figure \ref{fig:cylindricalcaseschematic}}
    \nomenclature[]{$(u,w)$}{melt velocity components in ($r,z$) directions}
    \nomenclature[G]{$\mu$}{dynamic viscosity}
    \nomenclature[G]{$\epsilon$}{aspect ratio of the characteristic melt film thickness $\delta_0$ and the characteristic length $R$}
    \nomenclature[G]{$\nu$}{kinematic viscosity}
    \nomenclature[G]{$\lambda$}{thermal conductivity}
    \nomenclature[G]{$\alpha$}{thermal diffusivity}
    \nomenclature[G]{$\Gamma$}{boundary}
    \nomenclature[G]{$\Omega$}{domain}
    \nomenclature[G]{$(\xi,\eta)$}{coordinates for the transformed energy equation}
    \nomenclature[G]{$\delta$}{melt film thickness}
    \nomenclature[G]{$\delta_0$}{characteristic melt film thickness}
    \nomenclature[G]{$\rho$}{density}
    \nomenclature[]{$k$}{iteration step}
    \nomenclature[U]{$\Box_a$}{analytical solution}
    \nomenclature[U]{$\Box_L$}{liquid state}
    \nomenclature[U]{$\Box_f$}{reference solution}
    \nomenclature[U]{$\Box_S$}{solid state}
    \nomenclature[U]{$\Box_i$}{iteration index}
    \nomenclature[U]{$\Box_w$}{heat source surface}
    \nomenclature[U]{$\Box_e$}{at the lateral outflow of the melt film, i.e. $r=\pm R$}
    \nomenclature[U]{$\Box_m$}{melting point of PCM}
    \nomenclature[X]{$\tilde{\Box}$}{dimensionless variable}
    \nomenclature[X]{$\Delta{\Box}$}{grid spacing}
    \nomenclature[X]{$d\Box/d r=\Box'$}{derivative with respect to $r$}
    \nomenclature[X]{$\partial \Box/\partial \Box$}{partial derivative}
    
\newcommand{\nomenclheader}[1]{\item[\hspace*{-\itemindent}#1]}
\renewcommand\nomgroup[1]{%
  \ifthenelse{\equal{#1}{A}}{%
   \nomenclheader{\textbf{Acronyms}}}{%                   A - Acronyms
    \ifthenelse{\equal{#1}{R}}{%
     \nomenclheader{\textbf{Roman Symbols}}}{%            R - Roman
      \ifthenelse{\equal{#1}{G}}{%
        \nomenclheader{\textbf{Greek Symbols}}}{%          G - Greek
          \ifthenelse{\equal{#1}{S}}{%
           \nomenclheader{\textbf{Superscripts}}}{{%      S - Superscripts
        \ifthenelse{\equal{#1}{U}}{%
         \nomenclheader{\textbf{Subscripts}}}{{%           U - Subscripts
        \ifthenelse{\equal{#1}{X}}{%
         \nomenclheader{\textbf{Other Symbols}}}%          X - Other Symbols
                       {{}}}}}}}}}}

\newcommand\caseA{cylindrical heat source}
\newcommand\CaseA{Cylindrical heat source}
\newcommand\caseB{rectangular heat source}
\newcommand\CaseB{Rectangular heat source}

\biboptions{square}

\makeatletter
\def\ps@pprintTitle{%
	\let\@oddhead\@empty
	\let\@evenhead\@empty
	\def\@oddfoot{}%
	\let\@evenfoot\@oddfoot}
\makeatother

\begin{document}

	\begin{frontmatter}
		
		
		
		%% optional Line numbering
				
		\title{A Lagrangian approach to modeling heat flux driven close-contact melting}
		
		\author[aRWTH]{K.~Sch\"uller\corref{cor1}}
		\ead{schueller@aices.rwth-aachen.de}
		\author[aRWTH]{J.~Kowalski}
		%\author[aRWTH]{J.~Subodh Rao}
		
		\cortext[cor1]{Corresponding author}
		
		\address[aRWTH]{Aachen Institute for Advanced Study in Computational Engineering Science, RWTH Aachen University, Schinkelstr. 2, 52062 Aachen, Germany}
		%\linenumbers
		\begin{abstract}
Close-contact melting refers to the process of a heat source melting its way into a phase-change material. Of special interest is the close-contact melting velocity, or more specifically the relative velocity between the heat source and the phase-change material. In this work, we present a novel numerical approach to simulate quasi-steady, heat flux driven close-contact melting. It extends existing approaches found in the literature, and, for the first time, allows to study the impact of a spatially varying heat flux distribution. We will start by deriving the governing equations in a Lagrangian reference frame fixed to the heat source. Exploiting the narrowness of the melt film enables us to reduce the momentum balance to the Reynolds equation, which is coupled to the energy balance via the velocity field. We particularize our derivation for two simple, yet technically relevant geometries, namely a 3d circular disc and a 2d planar heat source. An iterative solution procedure for the coupled system is described in detail and discussed on the basis of a convergence study. Furthermore, we present an extension to allow for rotational melting modes. Various test cases demonstrate the proficiency of our method. In particular, we will utilize the method to assess the efficiency of the close-contact melting process and to quantify the model error introduced if convective losses are neglected. Finally, we will draw conclusions and present an outlook to future work.
\end{abstract}
		
		
	\end{frontmatter}
	%\linenumbers
	\section{Introduction}
Computational contact mechanics is a wide and vivid field of research and of great importance in many science and engineering areas \cite{wriggers2006computational}. Oftentimes, one aims at computationally determining the evolving forces between contacting bodies and interfaces, while deformations are either small or of secondary importance. However, there are also other contact situations, one example being the process of close-contact melting, which will be the content of this study.

\begin{figure}
\centering
\includegraphics[width=0.6\linewidth]{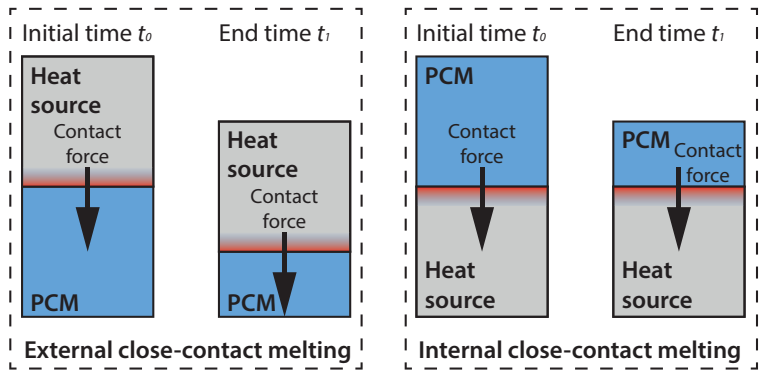}
\caption{External close-contact melting (left) and internal close-contact melting (right). Both external, and internal close-contact melting are governed by the same physical process.}
\label{fig:contactmelting}
\end{figure}

Close-contact melting occurs when a heated body and an initially solid phase change material (PCM) are in contact, and heat transfer across the contact area induces melting on the PCM side, see figure \ref{fig:contactmelting}. We will refer to the heated body as the heat source in the ongoing of this article. As soon as the PCM undergoes phase change, a narrow gap filled with liquid PCM material develops between the heat source and the solid PCM. This fact explains the naming close-contact melting, which emphasizes that the energy supplying heat source is physically separated from the solid-liquid phase interface by a micro-scale liquid filled gap. Consequently, the heat has to be transported through this micro-scale liquid melt film prior to inducing any phase change in the PCM.

Simultaneous melting and acting contact forces, e.g. due to gravity, leads to a relative motion of the heat source and the solid PCM. The literature distinguishes external close-contact melting, in which the PCM is stationary, and internal close-contact melting, in which the heat source is stationary \cite{yaojiang1999generalized}, see figure \ref{fig:contactmelting} for a schematic overview. The underlying physical process, however, is independent of the taken perspective and coincides in both cases. This is why we will not discriminate between internal and external close-contact melting in this publication.

Close-contact melting can be observed in various science and engineering applications, e.g. in latent heat storage systems that provide an effective way of storing thermal energy. In these systems the PCM can move freely within a temperature regulated chamber. Due to its own weight the PCM tends to sink down to the bottom wall at which close-contact melting can be observed. This can be exploited to increase the rate of heat transfer into the storage and hence to optimize the charging time \cite{kasibhatla2016numerical,gudibande2017numerical}. Other applications for close-contact melting are hot wire cutting manufacturing technology \cite{mayer2008close}, nuclear technology \cite{emerman1983stokes,chen2013study}, planetary protection \cite{lorenz2012thermal} as well as robotic ice exploration technologies based on melting probes for polar research and solar system research, e.g. \cite{kowalski2016navigation,zimmerman2001cryobot,cardell2004subsurface,erokhina2015technique}. 

Developing a computational model for robotic ice exploration technologies based on melting probes also serves as the major motivation for our current study. First patents \cite{adams1963nuclear,benson1968thermal}, experiments and theoretical studies \cite{shreve1962theory,aamot1967heat} regarding thermal melting probes for terrestrial applications date back to the 1960s. Since then, the potential of space exploration technologies based on melting probes has been assessed \cite{ulamec2007access}, and specific solar system mission concepts e.g. for exploring the Saturnian moon Enceladus \cite{2015:Konstantinidis}, have been proposed. In contrast to the rather sophisticated hardware concepts that have been developed and tested in recent years \cite{kowalski2016navigation}, only minor steps have been undergone to model the process \cite{schuller2016curvilinear,aamot1967heat,shreve1962theory,fomin1995contact}. In this study, we will concentrate on the close-contact melting processes in the vicinity of a robotic melting probe, as they are key to modeling the macro-scale dynamics of the probe. In this context, we will refer to the relative motion between the heat source and the PCM as the melting velocity at which the probe moves into the ice. It is important to note, however, that our work is not restricted to modeling ice exploration technologies, but can be applied to any close-contact melting situation. 

Modeling close-contact melting essentially boils down to modeling the micro-scale melt film that separates the heat source from the PCM. For heat sources of constant temperature (temperature driven close-contact melting), simple geometries (e.g. a sphere or a cylinder), as well as quasi-steady state conditions, it is possible to derive an analytical solution \cite{emerman1983stokes,moallemi1985melting} based upon the balance laws for mass, momentum and energy. Those solutions have been summarized in \cite{bejan1992single}. The results show consistently that in these cases the melting velocity $W$ scales with the contact force $F_\text{eff}$ according to $W\sim F_\text{eff}^{1/4}$. Recently, these results have been generalized to allow for rotational melting induced by an asymmetric surface temperature profile \cite{schuller2016curvilinear}.

Characteristic to the temperature driven close-contact melting process is that the melt film thickness asymptotically tends to zero for increasing contact forces. A decreasing melt film thickness, however, implies an increasing heat flux. This is unrealistic, as soon as the system is constraint by a maximal possible heat flux and motivates looking at heat flux driven close-contact melting in general, hence a Neumann type boundary condition for the temperature field at the heat source surface.

In contrast to quite a number of publications that address the temperature driven process, there are only few contributions that deal with heat flux driven close-contact melting \cite{moallemi1986analysis,zhao2008analysis}. 
In this publication, we propose a numerical model that generalizes existing approaches for computing heat flux driven close-contact melting. The strength of our approach is that we allow for a general heat flux distribution at the heat source surface, which can induce both translational as well as rotational modes of the melting motion. In particular, we will

\begin{enumerate}
	\item derive the mathematical model and governing equations including a scaling analysis for quasi-steady state close-contact melting (section \ref{sec:Derivation}). We will particularize our model for two types of heat source geometries, namely
	\begin{itemize}
	\item a cylindrical heat source for which the contact area is given as a circular disc, and
	\item a rectangular heat source for which the contact area is given as a 2d plane of infinite extend in cross-direction
	\end{itemize}
	\item describe and analyze an iterative numerical strategy to solve the close-contact melting system for translational melting velocity, melt film thickness profile, as well as temperature, velocity, and pressure field within the melt film (section \ref{sec:NumericalMethod})
	\item describe and analyze an extension to this iterative numerical strategy that accounts for an additional rotational melting modes (section \ref{sec:NumericalMethod})
	\item apply the models to analyze the efficiency of the close-contact melting processes in both translational and rotational melting motion (section \ref{sec:resultsAndDiscussion})
\end{enumerate}
	
	\begin{table*}[!t]
		\begin{framed}
			%\footnotesize
			\printnomenclature
		\end{framed}
	\end{table*}

	\section{Problem formulation}
\label{sec:Derivation}
\subsection{Physical model}
\begin{figure}
\centering
\includegraphics[width=1\linewidth]{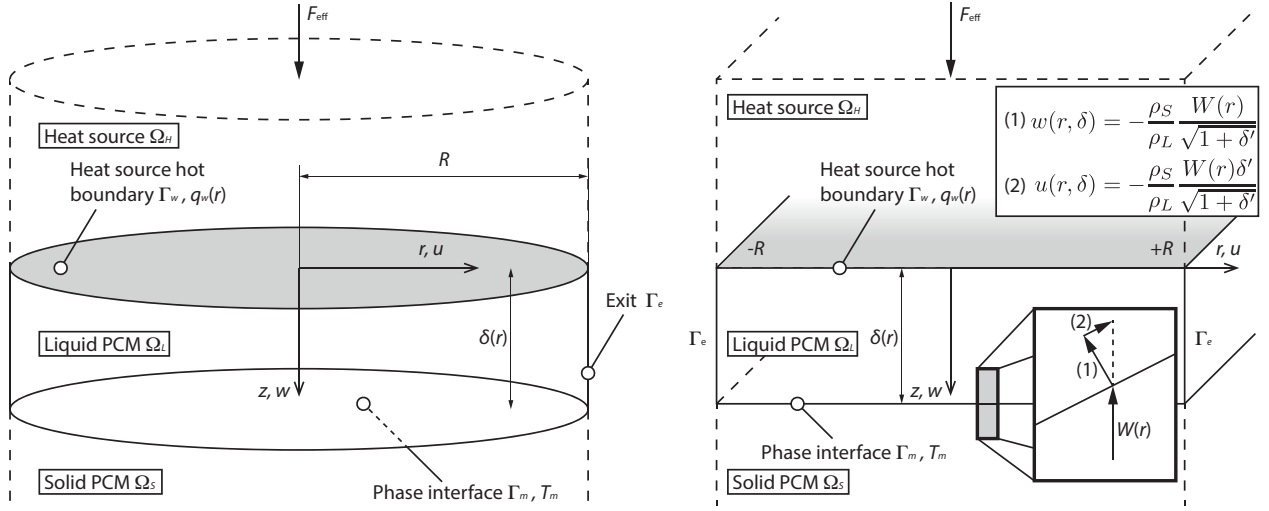}
\caption{Schematic of the physical problem for the two geometry settings: \CaseA\;$n=1$ (left) and \caseB\;$n=0$ (right). The figure on the right additionally zooms into the components of the inflow velocity $W(r)$ in normal and tangential direction evaluated at the solid-liquid interface. The latter holds for both cases.}
\label{fig:cylindricalcaseschematic}
\end{figure}

We look at close-contact melting for two specific geometry settings, namely a \caseA, as well as a \caseB. Both are sketched in figure \ref{fig:cylindricalcaseschematic}.
In each case, the heat source, given by $\Omega_H$, is in close-contact with a PCM that is denoted by the domain $\Omega_L$ and $\Omega_S$ for its liquid and solid phase, respectively. In the ongoing of this article, we will use the following naming conventions:
\begin{itemize}
	\item \CaseA\;($n=1$): A cylindrical heat source with the radius $R$ and a planar front face
	\item \CaseB\;($n=0$): A rectangular heat source  with a length of $2R$ for which the contact area is given as a 2d plane of infinite extent in cross-direction
\end{itemize}
%A cylindrical heat source with a planar front face and a cross-sectional radius R or a 2d planar heat source with a length of $2R$ is in contact with a PCM. 
%The existence of a liquid phase is the result of the applied heat flux $q_w$ that is acting on the working surface ($\Gamma_w$) and continuously cause phase change of the PCM. 
In each case, the coordinate system with axes $r$ and $z$ is fixed with respect to the heat source. It has its origin at the center of the heat source surface $\Gamma_w$, which will be denoted as working surface in the ongoing of this article. The heat flux $q_w$ is considered to be spatially varying, but constant in time. A constant force $F_\text{eff}$ is acting on the heat source, which causes squeezing of the melt film and, consequently, induces a velocity field in the liquid PCM, in which $u$ and $w$ are the velocity components in $r$- and $z$-direction, respectively. Due to mass conservation, there will be a continuous outflow of molten material through the lateral outflow boundary $\Gamma_e$. The location of the phase interface $\Gamma_m$ is given by the melt film profile $\delta$. Its temperature is equal to the PCM's melting temperature $T_m$. The melting velocity $W(r)$ is the relative motion between the working surface and the PCM.

We are interested in regimes that are characterized by slow melting velocities. Hence, we can safely assume instantaneous relaxation into a mechanical equilibrium, which means that quasi-steady state is attained soon after initiating the melting. This choice of scale is also supported by laboratory experiments \cite{moallemi1985experiments}.

\subsection{Mathematical model}
Starting from balance of mass, momentum and energy, we will simplify the governing equations by using the lubrication approximation for the melt film. This is a valid simplification, because of the very thin melt film compared to the length of the heat source (i.e. $\delta/R<<1$). Scaling analysis shows that inertia terms are negligible compared to the pressure gradient and also $\partial/\partial r^2<<\partial/\partial z^2$ as long as $\delta/R<<1$ and $(\delta/R) \text{Re}<<1$ \cite{moallemi1986experimental,hamrock2004fundamentals}.
According to \cite{moallemi1986experimental}, conservation of mass, momentum and energy for the incompressible Newtonian melt film within the liquid phase of the PCM $\Omega_L$ yields
\begin{align}
\label{eq:dimensionalMassEquation}
&\frac{1}{r^n}\frac{\partial}{\partial r}\left( r^n u \right)+\frac{\partial w}{\partial z}=0\\
\label{eq:dimensionalPressureEquation}
&\frac{d p}{dr}=\mu_L\frac{\partial^2 u}{\partial z^2},\qquad \frac{d p}{d z}=0\\
\label{eq:dimensionalEnergyEquation}
& \alpha_L\left[ \frac{1}{r^n}\frac{\partial}{\partial r}\left( r^n \frac{\partial T}{\partial r} \right)+\frac{\partial^2 T}{\partial z^2} \right]-\left( u\frac{\partial T}{\partial r}+w\frac{\partial T}{\partial z} \right)=0
\end{align}
in which $n=1$, $0<r\leq R$ for the \caseA\;and $n=0$, $-R\leq r\leq R$ for the \caseB. Parameters in this system are the thermal diffusivity $\alpha_L$ and the dynamic viscosity $\mu_L$ of the melt film. Under the stated assumptions, the pressure $p$ varies only in $r$-direction, whereas the temperature $T$ depends on both spatial coordinates.

At the working surface $\Gamma_w$ ($z=0$), we have no-slip, zero inflow and a spatially varying heat flux
\begin{align}
  \label{eq:boundaryCondu0}
	&u(x,0)=w(x,0)=0\\
	&\left. \frac{\partial T}{\partial z}\right|_{z=0}=-\frac{q_w(r)}{\lambda_L}
\end{align}
in which $\lambda_L$ is the thermal conductivity of the liquid PCM. Taking into account the density change due to phase change, the boundary conditions at the phase interface $\Gamma_m$ ($z=\delta$) are
\begin{align}
  \label{eq:boundaryCondu}
	&u(r,\delta)=-\frac{\rho_S}{\rho_L}W(r)\frac{\delta'}{\sqrt{1+\delta'^2}}\\
	&w(r,\delta)=-\frac{\rho_S}{\rho_L}\frac{W(r)}{\sqrt{1+\delta'^2}}\\
	&T(r,\delta)=T_m\\
	\label{eq:boundaryCondSte}
  &\left. \frac{\partial T}{\partial z}\right|_{z=\delta}=-\frac{\rho_S }{\lambda_L}\frac{W(r)\,h_m^*}{\sqrt{1+\delta'^2}} 
\end{align}
Here, $\rho_S$ and $\rho_L$ are the densities of the solid and liquid phase, and $\delta'=d\delta/d r$ is the spatial derivative of the melt film thickness. Boundary conditions \eqref{eq:boundaryCondu}--\eqref{eq:boundaryCondSte} hold for both geometries ($n=1$ and $n=0$). $h_{m}^*=h_m+c_{p,S}\left( T_m-T_S \right)$ denotes the reduced latent heat of melting that accounts for both, the latent heat of melting, as well as the amount of sensible heat required to raise the temperature of the solid PCM to the melting temperature. The melting velocity $W$ is the relative velocity between the heat source and the PCM. For straight melting it is constant and for curvilinear melting, it varies along the working surface $\Gamma_w$ as it is described in \cite{schuller2016curvilinear}.

The boundary conditions for the pressure and temperature fields depend on the geometry, and are given by
\begin{align}
	\text{n=1:}\qquad &\left.\frac{d p}{d r}\right|_{r=0}=0,\qquad p(R)=0\\
	&\left.\frac{\partial T}{\partial r}\right|_{r=0}=\left.\frac{\partial T}{\partial r}\right|_{r=R}=0\\
	\text{n=0:}\qquad &p(-R)=p(R)=0\\
	\label{eq:dimensionalBCs}
	&\left.\frac{\partial T}{\partial r}\right|_{r=-R}=\left.\frac{\partial T}{\partial r}\right|_{r=R}=0
\end{align}
Two length scales determine the system, namely the characteristic melt film thickness $\delta_0$ and the half-length of the heat source that is given by $R$. Their aspect ratio is denoted by $\epsilon=\delta_0/R$ and characterizes the narrowness of the melt film. We look at regimes in which $\epsilon$ is in the order of $10^{-3}$ or lower. The other scales follow naturally and we end up with the dimensionless variables
\begin{equation}
\begin{aligned}
	&r=\tilde{r} R,\quad z=\tilde{z} \delta_0,\quad u=\tilde{u}u_0,\quad w=\tilde{w}W_0,\qquad W=\tilde{W}W_0,\quad \delta=\tilde{\delta} \delta_0,\quad \epsilon=\delta_0/R,\\
	&T=\tilde{T}q_{w,\text{ref}}\delta_0/\lambda_L+T_m,\quad q_{w}=\tilde{q}_{w}\,q_{w,\text{ref}},\quad W_0=\frac{\tilde{W_0} \mu_L}{\rho_L R},\quad p=\tilde{p}W_0 R \mu_L/\delta_0^2,\\
	\label{eq:dimensionlessVariables}
	&Ste=\frac{q_{w,\text{ref}}\,c_{p,L}\,\delta_0}{\lambda_L\,h_m^*},\qquad\text{Re}=\rho_L W_0 \delta_0/\mu_L,\qquad u_0=W_0 R/\delta_0,\qquad Pe=W_0\delta_0/\alpha_L
\end{aligned}
\end{equation}
in which $\text{Ste}$, $\text{Re}$ and $\text{Pe}$ are the Stefan number, Reynolds number and Peclet number, respectively. Written in terms of the dimensionless variables \eqref{eq:dimensionlessVariables}, the balance laws \eqref{eq:dimensionalMassEquation}--\eqref{eq:dimensionalEnergyEquation} transform into
\begin{align}
  \label{eq:nonDimMass}
	&\frac{1}{\tilde{r}^{n}}\frac{\partial}{\partial \tilde{r}}\left( \tilde{r}^{n} \tilde{u} \right)+\frac{\partial \tilde{w}}{\partial \tilde{z}}=0\\
  \label{eq:nonDimMomentum}
        &\epsilon\frac{d \tilde{p}}{d \tilde{r}}=\frac{\partial^2 \tilde{u}}{\partial \tilde{z}^{2}}\\
        &\frac{d \tilde{p}}{d \tilde{z}}=0\\
        \label{eq:nonDimHeat}
	& \epsilon^2\frac{1}{\tilde{r}^{n}}\frac{\partial}{\partial \tilde{r}}\left( \tilde{r}^{n}\frac{\partial \tilde{T}}{\partial \tilde{r}} \right)+\frac{\partial^2 \tilde{T}}{\partial \tilde{z}^{2}} -Pe\left( \tilde{u}\frac{\partial \tilde{T}}{\partial \tilde{r}}+\tilde{w}\frac{\partial \tilde{T}}{\partial \tilde{z}}\right)=0
\end{align}
Substituting the dimensionless variables \eqref{eq:dimensionlessVariables} into the boundary conditions \eqref{eq:boundaryCondu0}--\eqref{eq:dimensionalBCs} yields their transformed version
\begin{align}
\label{eq:boundaryCondu0DimLess}
	&\tilde{u}(\tilde{r},0)=\tilde{w}(\tilde{r},0)=0\\
	\label{eq:heatFluxDimLessBC}
	&\left. \frac{\partial \tilde{T}}{\partial \tilde{z}}\right|_{\tilde{z}=0}=-\tilde{q}_w\\
  \label{eq:boundaryConduDimLess}
  &\tilde{u}(\tilde{r},\tilde{\delta})=-\frac{\rho_S}{\rho_L}\tilde{W}\frac{\epsilon^2\,\tilde{\delta}'}{\sqrt{1+(\epsilon\,\tilde{\delta}')^2}}\\
  \label{eq:BCInflowNonDim}
  &\tilde{w}(\tilde{r},\tilde{\delta})=-\frac{\rho_S}{\rho_L} \frac{\tilde{W}}{\sqrt{1+(\epsilon\,\tilde{\delta}')^2}} \\
  \label{eq:MeltingTemperatureTransformed}
  &\tilde{T}(\tilde{r},\tilde{\delta})=0\\
  \label{eq:StefanCondTransformed}
  &\left. \frac{\partial \tilde{T}}{\partial \tilde{z}}\right|_{\tilde{z}=\tilde{\delta}}=-\frac{\rho_S}{\rho_L}\frac{\mu_L }{R} \frac{\tilde{W}_0\,\tilde{W} h_m^*}{q_{w,\text{ref}}\sqrt{1+(\epsilon\,\tilde{\delta}')^2}}\\
  \label{eq:dimLessBCReynoldsEquationn1}
  	\text{n=1:}\qquad &\left.\frac{d \tilde{p}}{d \tilde{r}}\right|_{\tilde{r}=0}=0,\qquad \tilde{p}(1)=0,\qquad
  	\left.\frac{\partial \tilde{T}}{\partial \tilde{r}}\right|_{\tilde{r}=0}=\left.\frac{\partial \tilde{T}}{\partial \tilde{r}}\right|_{\tilde{r}=1}=0\\
  	\label{eq:dimLessBCReynoldsEquationn0}
  	\text{n=0:}\qquad &\tilde{p}(-1)=\tilde{p}(1)=0,\,\qquad
  	\left.\frac{\partial \tilde{T}}{\partial \tilde{r}}\right|_{\tilde{r}=-1}=\left.\frac{\partial \tilde{T}}{\partial \tilde{r}}\right|_{\tilde{r}=1}=0
\end{align}
Integrating the dimensionless momentum equation \eqref{eq:nonDimMomentum} twice with respect to $\tilde{z}$ and substituting in the boundary conditions \eqref{eq:boundaryCondu0DimLess} and \eqref{eq:boundaryConduDimLess} yields the dimensionless horizontal velocity component within the melt film
\begin{align}
  \label{eq:uNonDimEq}
  \tilde{u}=\frac{1}{2}\epsilon\frac{d \tilde{p}}{d \tilde{r}}\tilde{z}\left( \tilde{z}-\tilde{\delta} \right)-\frac{1}{\tilde{\delta}}\frac{\rho_S}{\rho_L}\tilde{W}\frac{\epsilon^2\,\tilde{\delta}'}{\sqrt{1+(\epsilon\,\tilde{\delta}')^2}} \tilde{z}
\end{align}
The vertical velocity component is found by substituting the horizontal component \eqref{eq:uNonDimEq} into conservation of mass \eqref{eq:nonDimMass}, which yields
\begin{align}
  \label{eq:velocityInNormalDirNondimensional}
  \tilde{w}=&\frac{\epsilon\tilde{z}^2}{12}\left[\frac{d^2 \tilde{p}}{d \tilde{r}^2}\left( 3\tilde{\delta}-2\tilde{z} \right)+\frac{1}{r^n}\frac{d \tilde{p}}{d \tilde {r}}\left( 3n\tilde{\delta}-2n \tilde{z} +3\tilde{r}^n\tilde{\delta}' \right)\right]\nonumber\\
  &+\frac{\epsilon^2\tilde{z} \rho_S\left[\tilde{\delta} ' \left(\epsilon ^2 \tilde{\delta} '^2+1\right) \left( \tilde{W} \left(n \delta  \tilde{r}^n-\tilde{r}^n  \tilde{\delta} '\right)+\tilde{r}^n \tilde{\delta} \tilde{W}'\right)+\tilde{r}^n\tilde{\delta}  \tilde{W} \tilde{\delta} ''\right]}{\tilde{r}^n \rho_L \tilde{\delta}^2\left( \epsilon^2\tilde{\delta}'^2+1 \right)^{3/2}}
\end{align}
Now, we evaluate the vertical component of the velocity \eqref{eq:velocityInNormalDirNondimensional} at $\tilde{z}=\tilde{\delta}$. Substituting in the inflow velocity \eqref{eq:BCInflowNonDim} yields an ordinary differential equation for the pressure along the melt film

\begin{align}
\label{eq:ReynoldsODEComplicated}
   &\epsilon\frac{d^2 \tilde{p}}{d \tilde{r}^2}+\epsilon\frac{1}{r^n}\frac{d \tilde{p}}{d \tilde {r}}\left( n+\frac{3\tilde{r}^n\tilde{\delta}'}{\tilde{\delta}} \right)
   +\frac{12\tilde{W}\rho_S}{\tilde{\delta}^3\rho_L\sqrt{\epsilon^2\tilde{\delta}'^2+1}}\nonumber\\
   &+\frac{12 \epsilon^2 \rho _S \left[\tilde{\delta}'\left(\epsilon ^2  \tilde{\delta} '^2+1\right) \left(\tilde{U} \left( n \tilde{\delta}+\tilde{r} ^n \tilde{\delta}'\right)+\tilde{r}^n \tilde{\delta}\tilde{W}'\right)+\tilde{r}^n   \tilde{W} \tilde{\delta} \tilde{\delta} ''\right]}{ \tilde{\delta}^4 \tilde{r}^n \rho _L  \left(\epsilon ^2  \tilde{\delta} '^2+1\right)^{3/2}} =0
\end{align}
A solution of this differential equation for the pressure $\tilde p$ requires to know the pressure $\tilde{p}_e$ at the lateral outflow boundary $\Gamma_e$.

For most close-contact melting problems, the ratio of the characteristic melt film thickness and the characteristic length is very small, i.e. $\epsilon=\delta_0/R<<1$. Equations \eqref{eq:uNonDimEq}--\eqref{eq:ReynoldsODEComplicated} can hence be simplified by neglecting terms that scale with $\epsilon^2$. This yields the final equations for pressure and velocity field within the melt film
\begin{align}
\label{eq:finalReynoldsEquation}
&\frac{d^2 \tilde{p}}{d \tilde{r}^2}+\frac{1}{\tilde{r}^n}\frac{d\tilde{p}}{d \tilde{r}}\left( n+\frac{3\tilde{r}^n\tilde{\delta}'}{\tilde{\delta}} \right)+\frac{12 \tilde{W}\rho_S}{\epsilon \tilde{\delta}^3 \rho_L}=0\\
\label{eq:geschwFeldu}
&\tilde{u}=\frac{1}{2}\epsilon\frac{d \tilde{p}}{d \tilde{r}}\tilde{z}\left( \tilde{z}-\tilde{\delta} \right)\\
\label{eq:geschwFeldw}
&\tilde{w}=\frac{\epsilon\tilde{z}^2}{12}\left[\frac{d^2 \tilde{p}}{d \tilde{r}^2}\left( 3\tilde{\delta}-2\tilde{z} \right)+\frac{1}{r^n}\frac{d \tilde{p}}{d \tilde {r}}\left( 3n\tilde{\delta}-2n \tilde{z} +3\tilde{r}^n\tilde{\delta}' \right)\right]\
\end{align}
The pressure can hence be calculated for a given melt film profile, and further post-processed into the velocity field. Likewise, the hydrodynamic force acting on the working surface can be calculated as the surface integral of the pressure $\tilde{p}$ over the working surface $\Gamma_w$. It is given by
\begin{align}
\label{eq:pressureForceEq}
\tilde{F}_{\tilde{p}}=\left[n\left( 2 \pi -1 \right) + 1 \right] \int_{n-1}^1 \tilde{p}\,r^n\, d\tilde{r}
\end{align}
The contact pressure, on the other hand, that is exerted by the contact force $F_\text{eff}$ is given by
\begin{align}
\label{eq:pEffEq}
	%p_\text{eff}=\frac{F_\text{eff}}{\left[n\left(\pi\,R-2\right)+2\right]R}
	p_\text{eff}=\frac{F_\text{eff}}{A},
\end{align}
in which $A=\left[n\left(\pi\,R-2\right)+2\right]R$ denotes the area of the working surface. Note that the physical dimension of $F_\text{eff}$ is either $\text{N}$ for the \caseA\;($n=1$) or $\text{N/m}$ for the \caseB\;($n=0$). Correspondingly, $A$ has the dimension $\text{m}^2$ for the \caseA\;($n=1$) or $\text{m}$ for the \caseB\;($n=0$).
Non-dimensionalizing the contact pressure \eqref{eq:pEffEq} according to \eqref{eq:dimensionlessVariables}, and multiplying with a reference area $\tilde A=\left[n\left(\pi-2\right)+2\right]$ yields the non-dimensional contact force
\begin{align}
\label{eq:nonDimExertedForce}
\tilde{F}_\text{eff}= \frac{ F_\text{eff}\epsilon^2}{ W_0\,\mu_L}\frac{\left[n\left(\pi-2\right)+2\right]}{\left[n\left(\pi\,R-2\right)+2\right]}
\end{align}
According to Newton's third law, the hydrodynamic force of the melt film \eqref{eq:pressureForceEq} must balance the exerted force \eqref{eq:nonDimExertedForce}. Equating both, substituting in the melting velocity scaling allows to solve for $\tilde{W}_0$:
\begin{align}
\label{eq:meltingVelocityEquation}
\tilde{W}_0=\epsilon^2\frac{F_\text{eff}\,R\,\rho_L}{\,\mu_L^2\,\tilde{F}_{p}}\frac{\left[n\left(\pi-2\right)+2\right]}{\left[n\left(\pi\,R-2\right)+2\right]}
\end{align}
The complete quasi-steady close-contact melting process can now be determined by solving the Reynolds equation \eqref{eq:finalReynoldsEquation} for the pressure field and the energy equation \eqref{eq:nonDimHeat} for the temperature field. Both are coupled by the velocity field \eqref{eq:geschwFeldu} and \eqref{eq:geschwFeldw}, and implicitly depend on the melt film thickness $\tilde{\delta}$ as well as the melting velocity $\tilde{W}_0$. Closure for the latter is provided by \eqref{eq:meltingVelocityEquation} and the Stefan condition \eqref{eq:StefanCondTransformed}.

\section{Numerical method}
\label{sec:NumericalMethod}
In this section we will describe the discretization of equations \eqref{eq:finalReynoldsEquation} and \eqref{eq:nonDimHeat}. Generally, we use structured grids with equidistant nodes, in which $i$ and $j$ denote the node index in $x$- and $z$-direction (or $\xi$- and $\eta$-direction in transformed coordinates), respectively. After that, we formulate an iterative procedure for the melt film thickness. Finally, we will present the global solution algorithm for the coupled problem for straight, as well as for curvilinear melting.
\subsection{Discretization of the Reynolds equation}
The dimensionless Reynolds equation \eqref{eq:finalReynoldsEquation} depends on $r$ only. It can be solved for an initial guess on the melt film thickness and boundary conditions \eqref{eq:dimLessBCReynoldsEquationn1} and \eqref{eq:dimLessBCReynoldsEquationn0} using standard 1d finite difference discretization. In this work, we use second order central differences for all terms, which yields
\begin{align}
\label{eq:discretizedReynolds}
-\frac{2}{(\Delta \tilde{r})^2}p_i+p_{i+1}\left[ \frac{1}{(\Delta \tilde{r})^2}+\frac{1}{2\Delta\tilde{r}\,\tilde{r}_i^n}\left( n+3\tilde{r}_i^n\frac{\tilde{\delta}_i'}{\tilde{\delta}_i} \right) \right]+p_{i-1}\left[ \frac{1}{(\Delta \tilde{r})^2}-\frac{1}{2\Delta\tilde{r}\,\tilde{r}_i^n}\left( n+3\tilde{r}_i^n\frac{\tilde{\delta}_i'}{\tilde{\delta}_i} \right) \right]-\frac{12\tilde{W}_i\,\rho_S}{\epsilon\,\tilde{\delta}_i^3\,\rho_L}=0
\end{align}
with
\begin{align}
\tilde{\delta}_i'=\frac{\tilde{\delta}_{i+1}-\tilde{\delta}_{i-1}}{2\Delta\tilde{r}}
\end{align}

\subsection{Discretization of the energy equation}
The energy equation \eqref{eq:nonDimHeat} is likewise discretized by finite differences. In order to carry out this task for variable melt film profiles $\tilde{\delta}(\tilde{r})$, we transform the melt film onto a rectangular grid via
\begin{figure}
	\centering
	\includegraphics[width=.6\linewidth]{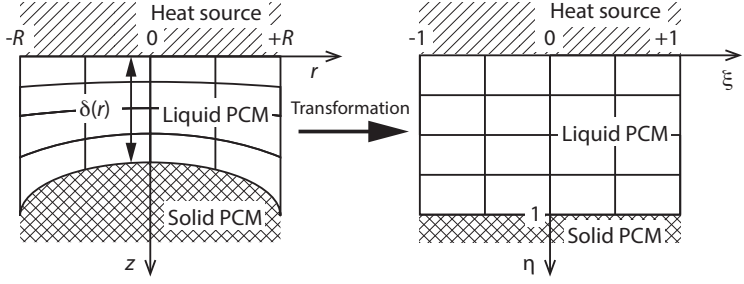}
	\caption{Sketch of the coordinate transformation. The liquid PCM domain $\Omega_L$ is located within curved grid lines for the physical domain (left), and rectangular grid lines for the computational domain (right).}
	\label{fig:transformsketch}
\end{figure}
\begin{align}
	\xi=\tilde{r},\qquad \eta=\frac{\tilde{z}}{\tilde{\delta}(\tilde{r})}
\end{align}
to map $\Omega_L$ onto a fixed rectangular domain in the computational ($\xi,\eta$) plane (see figure \ref{fig:transformsketch}) \cite{hamed1998numerical,hamed2000marangoni}. The differentials transform according to
\begin{align}
\label{eq:transformed1}
	\frac{\partial \tilde{T}}{\partial \tilde{r}}&=\frac{\partial \tilde{T}}{\partial \xi}\frac{\partial \xi}{\partial \tilde{r}}+\frac{\partial \tilde{T}}{\partial \eta}\frac{\partial \eta}{\partial r},\qquad
	\frac{\partial \tilde{T}}{\partial \tilde{z}}=\frac{\partial \tilde{T}}{\partial \xi}\frac{\partial \xi}{\partial \tilde{z}}+\frac{\partial \tilde{T}}{\partial \eta}\frac{\partial \eta}{\partial z}\\
\label{eq:transformed3}
	\frac{\partial^2 \tilde{T}}{\partial \tilde{r}^2}&=\frac{\partial^2 \tilde{T}}{\partial \xi^2}\left( \frac{\partial \xi}{\partial r} \right)^2+\frac{\partial^2 \tilde{T}}{\partial \eta^2}\left( \frac{\partial \eta}{\partial r} \right)^2+2\frac{\partial^2 \tilde{T}}{\partial \xi \partial \eta}\frac{\partial \xi}{\partial r}\frac{\partial \eta}{\partial r}+\frac{\partial \tilde{T}}{\partial \xi}\frac{\partial^2\xi}{\partial r^2}+\frac{\partial \tilde{T}}{\partial \eta}\frac{\partial^2\eta}{\partial r^2}\\
\label{eq:transformed4}
	\frac{\partial^2 \tilde{T}}{\partial \tilde{z}^2}&=\frac{\partial^2 \tilde{T}}{\partial \xi^2}\left( \frac{\partial \xi}{\partial z} \right)^2+\frac{\partial^2 \tilde{T}}{\partial \eta^2}\left( \frac{\partial \eta}{\partial z} \right)^2+2\frac{\partial^2 \tilde{T}}{\partial \xi \partial \eta}\frac{\partial \xi}{\partial z}\frac{\partial \eta}{\partial z}+\frac{\partial \tilde{T}}{\partial \xi}\frac{\partial^2\xi}{\partial z^2}+\frac{\partial \tilde{T}}{\partial \eta}\frac{\partial^2\eta}{\partial z^2}
\end{align} 

Substituting equations \eqref{eq:transformed1} to \eqref{eq:transformed4} into the energy equation \eqref{eq:nonDimHeat} yields
\begin{align}
	&\epsilon^2\left[ \frac{n}{\tilde{r}}\left( \frac{\partial \tilde{T}}{\partial \xi} -\frac{\eta}{\tilde{\delta}}\frac{d \tilde{\delta}}{d\xi}\frac{\partial \tilde{T}}{\partial \eta} \right)
	+ \frac{\partial^2\tilde{T}}{\partial \xi^2}+\frac{\partial^2 \tilde{T}}{\partial \eta^2}\left(\frac{\eta}{\tilde{\delta}}\frac{d\tilde{\delta}}{d\xi}\right)^2-2\frac{\eta}{\tilde{\delta}}\frac{d\tilde{\delta}}{d\xi}\frac{\partial^2 \tilde{T}}{\partial \xi \partial \eta}+\frac{\partial \tilde{T}}{\partial \eta}\frac{\eta}{\tilde{\delta}}\left( \frac{2}{\tilde{\delta}}\frac{d\tilde{\delta}}{d\xi}-\frac{d^2\tilde{\delta}}{d\xi^2} \right) \right]\nonumber\\
	&\qquad +\frac{1}{\tilde{\delta}^2}\frac{\partial^2 \tilde{T}}{\partial \eta^2} -Pe\left[ \tilde{u} \left( \frac{\partial \tilde{T}}{\partial \xi}-\frac{\eta}{\tilde{\delta}}\frac{d \tilde{\delta}}{d \xi}\frac{\partial \tilde{T}}{\partial \eta} \right)+\tilde{w}\frac{1}{\tilde{\delta}}\frac{\partial \tilde{T}}{\partial \eta} \right]=0
\end{align}

We again neglect terms that scale with $\epsilon^2$, which after some reordering yields the final energy equation
\begin{align}
\label{eq:finalHeatEquationNondimensionalized}
	\frac{\partial^2 \tilde{T}}{\partial \eta^2}-Pe\,\tilde{u}\,\tilde{\delta}^2 \frac{\partial \tilde{T}}{\partial \xi}
	-Pe\,\tilde{\delta}\left[\tilde{w}-\tilde{u}\,\eta\frac{d\tilde{\delta}}{d\xi} \right]\frac{\partial \tilde{T}}{\partial \eta}=0
\end{align}
Since the cross-derivatives vanish for $\epsilon^2<<1$, we can use a five-point finite difference stencil in order to discretize equation \eqref{eq:finalHeatEquationNondimensionalized}. For the diffusion terms we use second order central differences. Although the Peclet number is small for the cases that we are interested in, the derivative of the dimensionless melt film thickness and hence the combined term ($\text{Pe}\,\tilde{\delta}\,\tilde{u}\,\eta\,d\tilde{\delta}/d\xi$) may become large. Therefore, we choose to apply a first order upwind discretization for the convection terms in \eqref{eq:finalHeatEquationNondimensionalized}. The resulting discretized energy equation is
\begin{align}
\label{eq:discretizedEnergyEquation}
&T_{i,j}\left( \frac{-2}{\left(\Delta \eta\right)^2}+\frac{a_{i,j}^- - a_{i,j}^+}{\Delta \xi}+\frac{b_{i,j}^- - b_{i,j}^+}{\Delta \eta} \right)-T_{i+1,j}\frac{a_{i,j}^-}{\Delta \xi}+T_{i-1,j}\frac{a_{i,j}^+}{\Delta \xi}+T_{i,j+1}\left(\frac{1}{\left(\Delta \eta\right)^2}-\frac{b_{i,j}^-}{\Delta \eta}  \right)+T_{i,j-1}\left(\frac{1}{\left(\Delta \eta\right)^2}+\frac{b_{i,j}^+}{\Delta \eta}  \right)=0\\
&a_{i,j}^+=\max\left( Pe\,\tilde{u}_{i,j}\,\tilde{\delta}_i^2,0 \right),\quad a_{i,j}^-=\min\left( Pe\,\tilde{u}_{i,j}\,\tilde{\delta}_i^2,0 \right),\\
&b_{i,j}^+=\max\left( Pe\,\tilde{\delta}_i\left( \tilde{w}_{i,j}-\tilde{u}_{i,j}\,\tilde{\eta}_j\frac{\tilde{\delta}_{i+1}-\tilde{\delta}_{i-1}}{2\Delta \xi} \right),0 \right),\quad b_{i,j}^-=\min\left( Pe\,\tilde{\delta}_i\left( \tilde{w}_{i,j}-\tilde{u}_{i,j}\,\tilde{\eta}_j\frac{\tilde{\delta}_{i+1}-\tilde{\delta}_{i-1}}{2\Delta \xi} \right),0 \right)
\end{align}
Boundary conditions are given by the heat flux at the working surface $\Gamma_w$ \eqref{eq:heatFluxDimLessBC}, the Dirichlet condition at the phase interface $\Gamma_m$ \eqref{eq:MeltingTemperatureTransformed} and the geometry-specific conditions at the lateral outflow $\Gamma_e$ \eqref{eq:dimLessBCReynoldsEquationn1} and \eqref{eq:dimLessBCReynoldsEquationn0} for $n=1$ and $n=0$, respectively.

\subsection{Calculation of the melt film thickness}
We exploit the fact that the temperature field is overdetermined as it has to fulfill both the Dirichlet condition \eqref{eq:MeltingTemperatureTransformed} and the Stefan condition \eqref{eq:StefanCondTransformed}, which is of Neumann type. We will hence iterate into the correct melt film thickness by solving the energy equation \eqref{eq:finalHeatEquationNondimensionalized} with the Dirichlet boundary condition at the phase interface for an initially guessed melt film profile $\tilde{\delta}$. The computed temperature field is then evaluated for its heat flux at the phase interface according to
\begin{align}
\label{eq:heatFluxCalculationEquation}
	\left.\frac{\partial \tilde{T}}{\partial z}\right|_{z=\tilde{\delta}}\approx \frac{3 \tilde{T}_{i,n_z}-4\tilde{T}_{i,n_z-1}+\tilde{T}_{i,n_z-2}}{2 \tilde{\delta}_i \Delta \eta}
\end{align}
At the corner nodes ($i=1,j=n_z$ and $i=n_r,j=n_z$) we use quadratic interpolation of the neighboring heat flux values at the phase interface. The ratio of the heat flux, computed in iteration step $k$ at the phase interface, and the Stefan condition is then given by
\begin{align}
	s_i^k=\frac{q_{w,\text{ref}}\left(3 \tilde{T}_{i,n_z}^k-4\tilde{T}_{i,n_z-1}^k+\tilde{T}_{i,n_z-2}^k\right)}{2 \tilde{\delta}_i^k \Delta \eta \rho_S W_0^k \tilde{W}^k h_m^*}
\end{align}
$s_i^k$ is hence a measure for the deviation from the correct melt film thickness and can be used to update the melt film thickness according to
\begin{align}
\label{eq:meltFilmThicknessUpdate}
	\tilde{\delta}_i^{k+1}=\left[ 1+\sigma\left( s_i^k -1\right) \right] \tilde{\delta}_i^k
\end{align}
in which $0<\sigma<1$ is a relaxation factor that we use to stabilize the procedure.

\subsection{Summary of the solution algorithm}
\label{sec:solutionalg}
\subsubsection{Straight close-contact melting}
\label{sec:straightMeltingSolutionAlgorithm}
In the case of straight melting the melting velocity $\tilde{W}(\tilde{r})$ is constant and hence $d\tilde{W}(\tilde{r})/d\tilde{r}=0$. In order to fully determine the close-contact melting process for both geometries ($n=1$ and $n=0$), we apply the following iterative solution procedure:
\begin{enumerate}
	\item Initial guess for the melt film profile.
	\item\label{itm:reynoldsEquation} Solve the Reynolds equation \eqref{eq:discretizedReynolds} to obtain the pressure distribution.
	\item Calculate the melting velocity by using equation \eqref{eq:meltingVelocityEquation}.
	\item Solve the transformed energy equation \eqref{eq:discretizedEnergyEquation} to obtain the temperature distribution in the melt film.
	\item Calculate the heat flux at the phase interface using \eqref{eq:heatFluxCalculationEquation}.
	\item\label{itm:updateMeltFilmThickness} Update the melt film thickness using \eqref{eq:meltFilmThicknessUpdate} and go back to step \ref{itm:reynoldsEquation} if the absolute error of the melting velocity is not smaller than a certain tolerance.
\end{enumerate}

\subsubsection{Curvilinear close-contact melting}
We consider curvilinear melting only for geometry setting $n=0$. The general approach follows \cite{schuller2016curvilinear} that first considered rotational melting modes. First, we assume that the exerted force is located at the center ($\tilde{r}=0$) of the heat source and acts normal to the working surface. In combination with an asymmetric heat flux profile, this will imply a rotating motion and hence the velocity profile $\tilde{W}(\tilde{r})$ varies along the heat source.  For quasi-steady close-contact melting, the resulting parametrized dimensionless melting velocity profile yields
\begin{align}
\label{eq:meltingVelocityProfile}
	\tilde{W}(\tilde{r})=1-\frac{\tilde{r}}{\tilde{r}_c}
\end{align}
in which $\tilde{r}_c=R\,r_c$ is the dimensionless curve radius given by the half-length of the heat source $R$ times the dimensional curve radius $r_c$.

In order to close the model, one additional equation is required. Due to constant angular velocity, the torque around the center of the heat source must be zero. It follows that the first moment of the hydrodynamic pressure has to vanish
\begin{align}
\label{eq:torqueEquation}
	\int_{-1}^1 \tilde{p}\,\tilde{r}\,dr=0
\end{align}
We can use this to extend the solution algorithm described in \ref{sec:straightMeltingSolutionAlgorithm} to curvilinear close-contact melting.
All in all, our solution algorithm for curvilinear close-contact melting of the \caseB\;is given by:
\begin{enumerate}
	\item Initial guess for both the melt film profile, and curve radius.
	\item\label{itm:reynoldsEquation2dplanar} Solve the Reynolds equation \eqref{eq:discretizedReynolds} together with \eqref{eq:meltingVelocityProfile} to obtain the pressure distribution.
	\item As long as the torque around the heat source center is not smaller than a certain tolerance, i.e. equation \eqref{eq:torqueEquation} is not fulfilled, update the curve radius (e.g. by linear extrapolation) and go back to step \ref{itm:reynoldsEquation2dplanar}.
	\item Calculate the melting velocity by using equation \eqref{eq:meltingVelocityEquation}.
	\item Solve the transformed energy equation \eqref{eq:discretizedEnergyEquation} to get the temperature distribution in the melt film.
	\item Calculate the heat flux at the phase interface using \eqref{eq:heatFluxCalculationEquation}.
	\item Update the melt film thickness using \eqref{eq:meltFilmThicknessUpdate} and go back to step \ref{itm:reynoldsEquation2dplanar} if the absolute error of the melting velocity is not smaller than a certain tolerance.
\end{enumerate}

	\subsection{Convergence analysis}
\label{sec:convergence}
\begin{figure}
\centering
\includegraphics[width=0.7\linewidth]{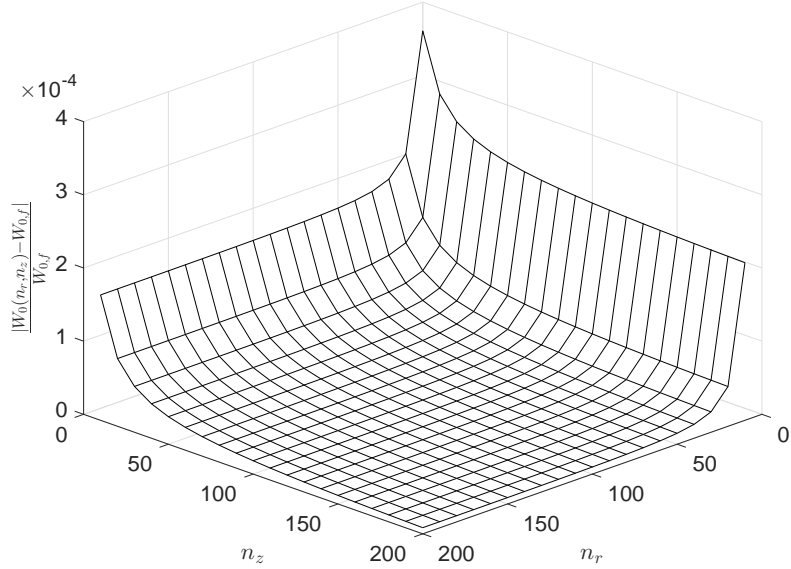}
\caption{Relative error of the melting velocity $W_0$ with respect to a reference solution $W_{0,f}$ computed on a fine mesh ($n_r=n_z=1000$). The plot shows our results for straight melting of the \caseA\;($R=0.1\,\text{m}$) with an exerted force of $1000\,\text{N}$ and a constant heat flux distribution of $100\,\text{kW/m}^2$ in standard $0\,^\circ\text{C}$ water-ice.}
\label{fig:relativeerrorgridrefinement}
\end{figure}

At first, we carry out a convergence study, in which we compare the melting velocity $W_0$ computed at mesh resolution $n_r$ and $n_z$ with a reference solution computed on a fine mesh $W_{0,f}$ ($n_r=n_z=1000$). Results are shown in figure \ref{fig:relativeerrorgridrefinement} which plots the relative error $|W_0-W_{0,f}|/W_{0,f}$ as a function of horizontal resolution $n_r$ and vertical resolution $n_z$. Even for a coarse mesh with $n_r=n_z=10$, the relative error is below $3.8\times 10^{-4}$. As expected, the error monotonically decreases as the number of nodes is increased. It is also evident that the relative error is more sensitive to changes in the horizontal resolution $n_r$ than to changes in the vertical resolution $n_z$. In other words, for the accuracy of the result the resolution along the melt film is more important than the resolution across the melt film. The latter is only felt by the temperature equation and not by the Reynolds equation \eqref{eq:finalReynoldsEquation}, which is formulated in a depth-integrated manner.

\begin{figure}
\centering
\includegraphics[width=0.8\linewidth]{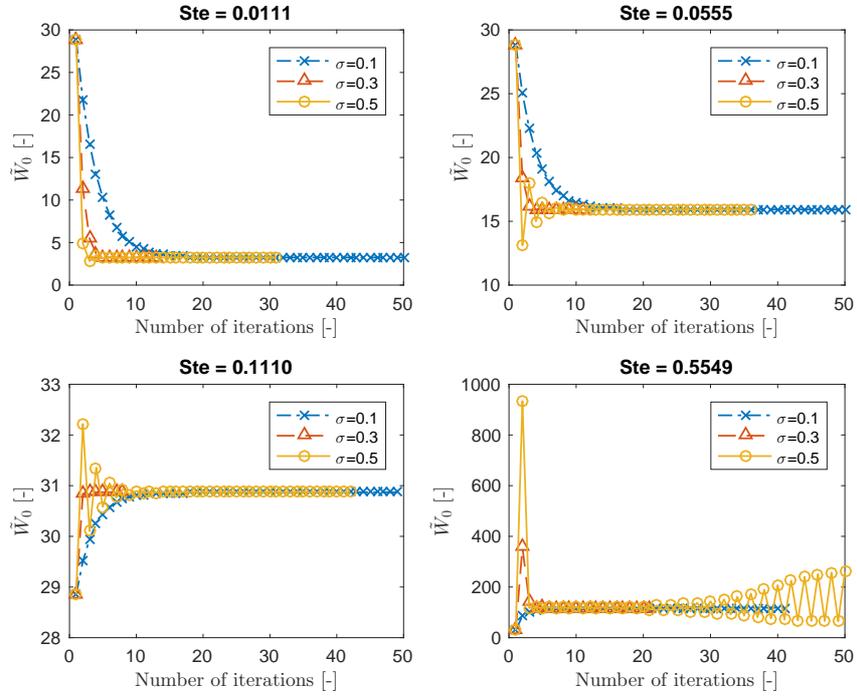}
\caption{Constant heat flux distribution: Convergence rate for different Stefan numbers and relaxation factors. For a Stefan number of 0.0111, a relaxation factor of 0.5 assures convergence, whereas for a Stefan number of 0.5549 the relaxation factor has to be chosen smaller.}
\label{fig:convergencerate}
\end{figure}

\begin{figure}
\centering
\includegraphics[width=0.8\linewidth]{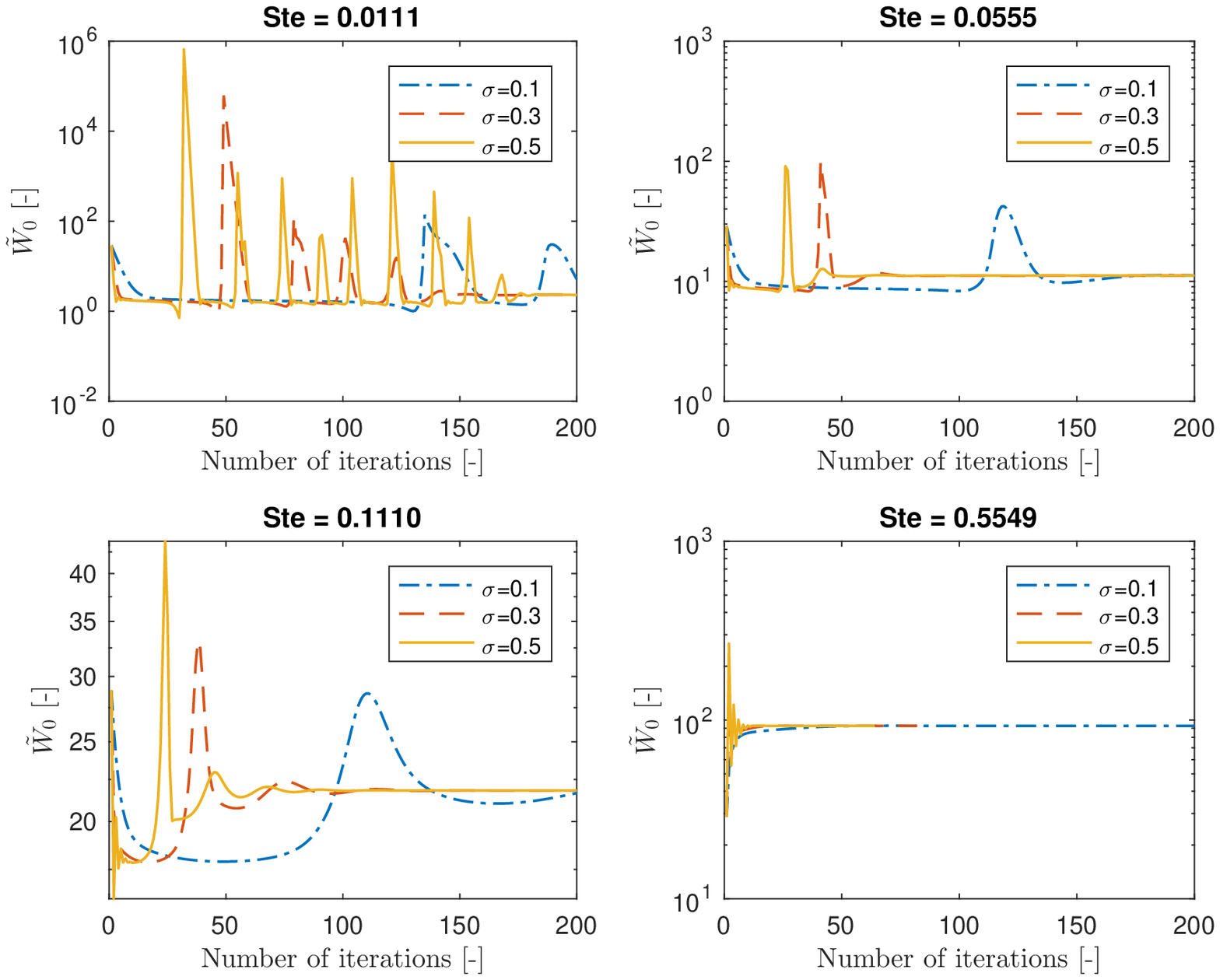}
\caption{Varying heat flux distribution: Convergence rate for different Stefan numbers and relaxation factors. Although the amplitudes early on in the iterative procedure are higher, convergence can be recovered for an appropriately small choice of the relaxation factor.}
\label{fig:convergencerateinhomogeneoushfprofile}
\end{figure}

Figure \ref{fig:convergencerate} and \ref{fig:convergencerateinhomogeneoushfprofile} show the empirical convergence rates for different values of the relaxation factor $\sigma$ at four different Stefan numbers. The Stefan number indicates the relative importance of sensible heat and the latent heat. A high Stefan number hence corresponds to a high input of heat at the working surface, whereas a low Stefan number represents a regime, in which the required latent heat for melting is considerably higher than the available heat input. For the figures we assumed an exerted force of $1000\,\text{N}$ and a convergence threshold of $10^{-8}$ on the dimensionless velocity. Furthermore, we used $n_r=40$ and $n_z=20$. In the first series of plots, see figure \ref{fig:convergencerate}, we assumed a constant heat flux distribution. As expected, the solution converges faster as we increase the relaxation factor $\sigma$. However, for a Stefan number of 0.5549, we found that a relaxation factor of $\sigma=0.5$ yields no convergence at all. Convergence can be recovered by lowering $\sigma$.

The convergence for a linear heat flux distribution of $\tilde{q}_w(\tilde{r})=2/3(\tilde{r}+1)$ can be seen in figure \ref{fig:convergencerateinhomogeneoushfprofile}. This time, the oscillations have higher amplitudes. This indicates that not only the mean melt film thickness, but its shape has a huge impact on the melting velocity.
Again, convergence can be achieved as we lower the relaxation factor. 
All in all, it can be stated that for low Stefan numbers, a higher relaxation factor can be used aiming for fast convergence, while for high Stefan numbers a smaller value should be used in order to assure convergence. 
A further analysis, however, that addresses the best choice for a relaxation factor, which provides a fast and guaranteed convergence is beyond the scope of this work. For the test cases considered in the ongoing of this study, computational cost is not an issue, which is why we use a value of $\sigma\leq0.1$ that always results in a stable scheme.


\section{Results and discussion}
\label{sec:resultsAndDiscussion}
For the test cases considered in the following section, we use the parameters that are summarized in table \ref{tb:parameters}. The thermo-physical parameters we used to create the solutions within this work are those of standard water-ice at $0\,^\circ \text{C}$.

\begin{table}
	\centering
	\caption{Default parameters used in this work.}
\begin{tabular}{lllllll}
	\hline 
	Parameter & Value & Unit & \hspace{1em} & Parameter & Value & Unit \\ 
	\hline 
	$h_m$ & 333700 & J/kg & & $\lambda_L$ & 0.57 & W/(m\,K) \\ 
	
	$\rho_L$ & 1000 & kg/m$^3$ && $\rho_S$ & 920 & kg/m$^3$ \\
	
	$c_{p,L}$ & 4222.2 & J/(kg\,m) && $c_{p,S}$ & 2049.41 & J/(kg\,m) \\
	
	$\mu_L$ & 0.001 & N\,s/m$^2$ && $T_S$ & 0 & $^\circ$C \\
	
	$T_m$ & 0 & $^\circ$C && R & 0.1 & m \\
	$F_\text{eff}$ & 1000 & N && $\hat{F}_\text{eff}$  & 1000 & N/m \\
	$n_r$ & 40 & - && $n_z$  & 20 & - \\
	\hline 
\end{tabular} 
\label{tb:parameters}
\end{table}

\subsection{Processes in the melt film}
\begin{figure}
	\hspace{-2em}
	\begin{subfigure}{0.55\textwidth}
		\includegraphics[width=\linewidth]{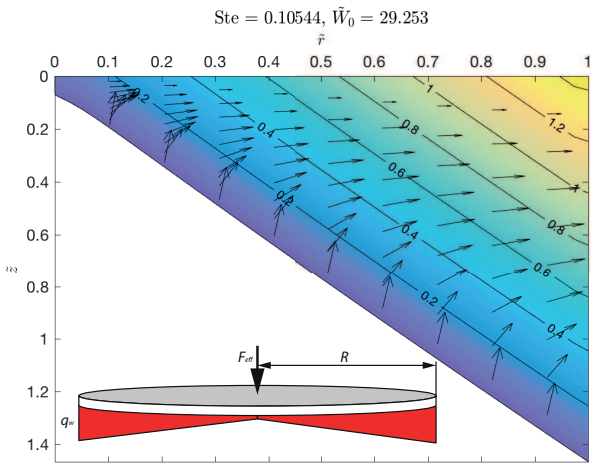}
		\caption{$a=-0.1$}
		\label{fig:velocityPlot1}
	\end{subfigure}
	\hspace{-2em}
	\begin{subfigure}{0.55\textwidth}
		\includegraphics[width=\linewidth]{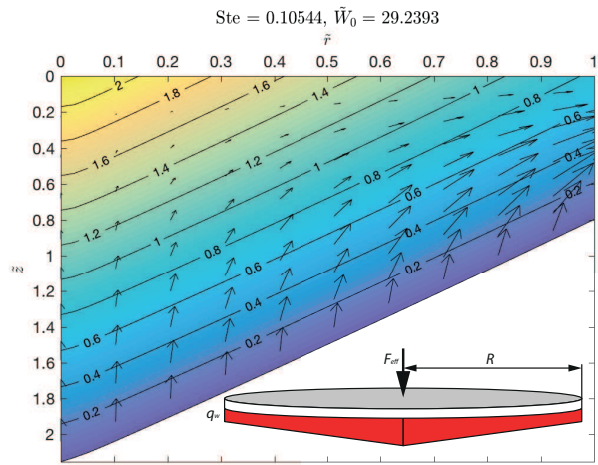}
		\caption{$a=0.1$}
		\label{fig:velocityPlot2}
	\end{subfigure}
\begin{center}
			\begin{subfigure}{0.55\textwidth}
			\includegraphics[width=\linewidth]{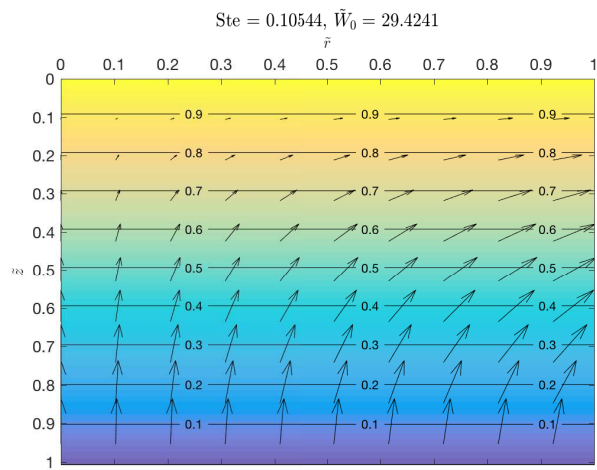}
			\caption{$a=0$}
			\label{fig:velocityPlotConstant}
		\end{subfigure}
\end{center}
	\caption{Temperature and velocity fields assuming a \caseA\;(n=1) for three different heat flux profiles : (a) a negative heat flux gradient, (b) a positive heat flux gradient, and (c) a constant heat flux profile, see also equation \eqref{eq:linearHeatFluxProfiles}.} \label{fig:velocityPlot}
\end{figure}

Figure \ref{fig:velocityPlot} considers the geometric setting that corresponds to the \caseA\;($n=1$) and shows velocity and temperature fields within the melt film for three different situations: (a) a negative heat flux gradient, (b) a positive heat flux gradient, and (c) a constant heat flux profile, given by
\begin{align}
	\label{eq:linearHeatFluxProfiles}
	\tilde{q}_w=\frac{1-a\,\tilde{r}}{1-a/2}
\end{align}
The parameters for figure \ref{fig:velocityPlot1}, \ref{fig:velocityPlot2} and \ref{fig:velocityPlotConstant} are chosen to be $a=-0.1$, $a=0.1$ and $a=0$, respectively. The Stefan number is the same for all three profiles.

Our first finding is that the melt film thickness of the constant heat flux profile (figure \ref{fig:velocityPlotConstant}) is also constant. This corresponds to experimental observations as well as analytical results \cite{moallemi1986experimental,yoo2000analytical}.\\ 
Moreover, we find that the computed velocities are almost the same in all three cases. Their heat flow rate, however, given by the heat flux surface integral
\begin{align}
\label{eq:dimHeatFlowRate}
	\tilde{Q}=2\pi\int_{0}^{1} \tilde{q}_w\,\tilde{r}\,d\tilde{r}=-\frac{1}{3}\pi\left( 5a-3 \right)
\end{align}
differs substantially. By evaluating the dimensionless heat flow rate \eqref{eq:dimHeatFlowRate} for all three situations, we find that $\tilde{Q}(a=0)=6/7\tilde{Q}(a=-0.1)=6/5\tilde{Q}(a=0.1)$.
This means that by concentrating the heat near the heat source's center, we can achieve the same melting velocity with $30\,\%$ less input heat. Hence, case (b) is most effective in terms of melting velocity. In this case, both the sensible heat is less, and the melt film thickness at the lateral outflow is small. This is contrasted by case (a), in which more heat leaves the system without contributing to the actual melting. Consequently, convective losses are higher.

\subsection{Straight close-contact melting versus optimal melting}
We are now able to assess the efficiency of straight close-contact melting. We do so by comparing our computational model, which considers convective losses, with an optimal melting velocity given by
\begin{align}
\label{eq:simpleEnergyBalanceEquation}
	\tilde{W}_{0,a}=\frac{q_{w}\,\rho_L\,R}{\mu_L\,\rho_S\, h_m^*}
\end{align}
It is worth mentioning that the efficiency can also be reinterpreted as the model error one introduces, when computing the melting velocity of a physical close-contact melting process by means of a simple energy balance, e.g. \cite{aamot1967heat}.

\begin{figure}
	\begin{subfigure}{0.5\textwidth}
		\includegraphics[width=\linewidth]{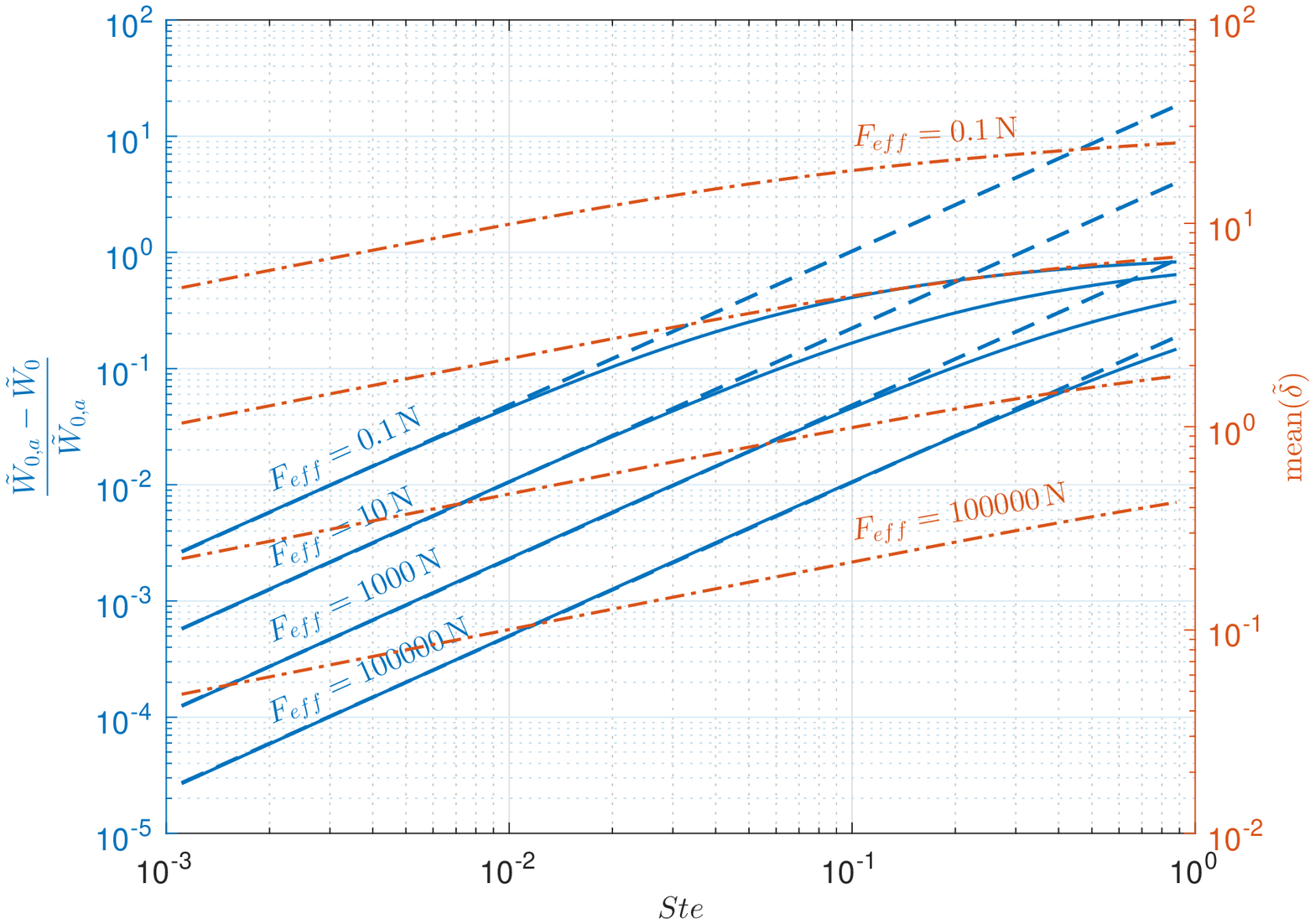}
		\caption{Axi-symmetric heat source (n=1)}
		\label{fig:curvefittingplot}
	\end{subfigure}
	\hspace{0em}
	\begin{subfigure}{0.5\textwidth}
		\includegraphics[width=\linewidth]{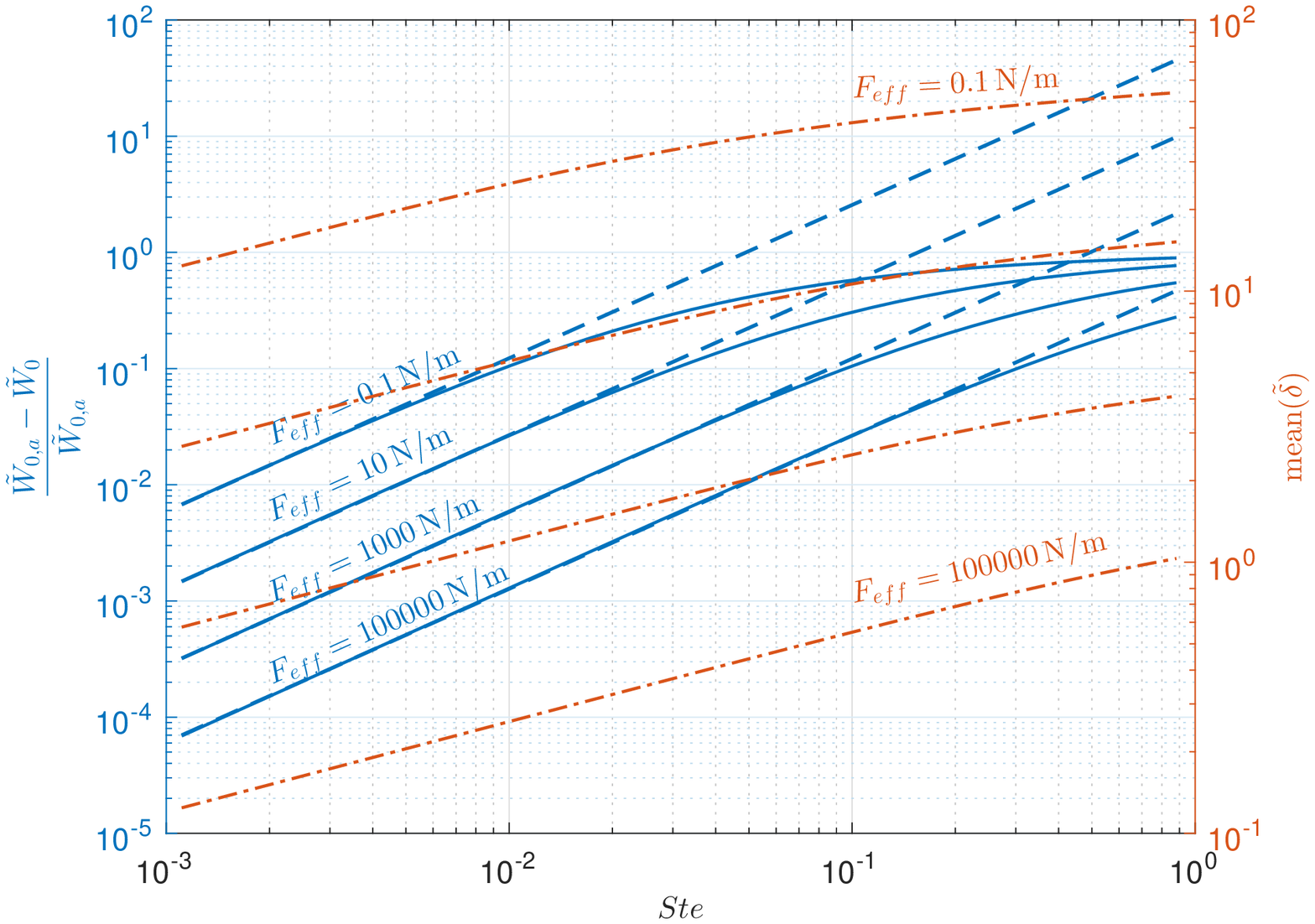}
		\caption{2d planar heat source (n=0)}
		\label{fig:curvefittingplot2d}
	\end{subfigure}
	\caption{Relative difference between the melting velocity determined with the computational close-contact model $\tilde{W}_0$, and the optimal melting velocity $\tilde{W}_{0,a}$ over the Stefan number for different exerted forces (solid lines); and the corresponding values of an approximation obtained by curve fitting for low Stefan numbers (linear dashed lines). The approximation is $(\tilde{W}_{0,a}-\tilde{W}_{0})/\tilde{W}_{0,a}\approx P_1 F_\text{eff}^{P_2}Ste^{P_3}$ with $P_1\approx10.06$, $P_2\approx-0.33$ and $P_3\approx1.32$ for $n=1$ and $P_1\approx24.92$, $P_2\approx-0.33$ and $P_3\approx1.32$ for $n=0$.} \label{fig:curvefittingplotBoth}
\end{figure}

Figure \ref{fig:curvefittingplotBoth} shows the relative difference between the close-contact melting velocity $\tilde{W}_0$ and the optimal melting velocity $\tilde{W}_{0,a}$ obtained by \eqref{eq:simpleEnergyBalanceEquation} as a function of the Stefan number, and for different exerted forces. The figure also includes the mean melt film thickness. All results are shown for both geometric settings. Since the computational close-contact melting model takes convective losses into account, its resulting melting velocity $\tilde{W}_0$ is always smaller than the optimal melting velocity. The difference becomes smallest for high forces. This is also what we expected, because the melt film thickness decreases with increasing forces. For a decreasing melt film thickness the transport of heat is more efficient and less prone to convective losses. The same tendency, though less dominant, can be observed as we lower the Stefan number.\\
We observed that for low values of the Stefan number and high contact forces, we can fit the relative distance between computed close-contact melting velocity and optimal melting velocity according to 
\begin{align}
\label{eq:approximationVelocityDifference}
	\frac{\tilde{W}_{0,a}-\tilde{W}_{0}}{\tilde{W}_{0,a}}= P_1 F_\text{eff}^{P_2}Ste^{P_3}
\end{align}
in which $P_1\approx10.06$, $P_2\approx-0.33$ and $P_3\approx1.32$ for $n=1$ and $P_1\approx24.92$, $P_2\approx-0.33$ and $P_3\approx1.32$ for $n=0$. Interestingly, $P_2$ and $P_3$ stay constant for both geometric settings. A further sensitivity analysis also shows that it is only $P_1$ that depends on the thermo-physical parameters. This could be exploited in the future to construct a simple analytical relation for assessing the close-contact melting velocities.

\subsection{Curvilinear close-contact melting versus optimal melting}
\begin{figure}
\centering
\includegraphics[width=0.55\linewidth]{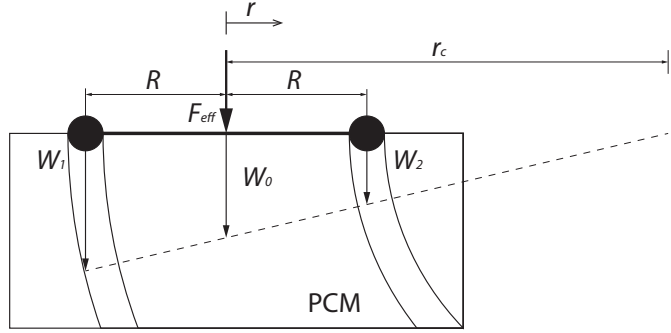}
\caption{Schematic of the analytical curvilinear close-contact melting model with two point heat sources.}
\label{fig:approximation}
\end{figure}

Again, we consider an ideal case without convective losses as our reference. Since the optimal curvilinear melting reference is less intuitive, we devote a first paragraph to constructing it. As in the previous section, we assume optimal melting according to \eqref{eq:simpleEnergyBalanceEquation}. Only this time, we consider two point heat sources, which are connected by a fully permeable rod so that their relative distance is fixed but the motion of the system is not influenced by the rod. A sketch of the situation is shown in figure \ref{fig:approximation}. The required heat flux for the point heat source $i$ follows from \eqref{eq:simpleEnergyBalanceEquation}
\begin{align}
	q_{w,i}=W_i\,\rho_S\, h_m^*
\end{align}
The velocities $W_{0,i}$ can be substituted by the linear velocity profile \eqref{eq:meltingVelocityProfile}, which yields for the left (index 1) and right (index 2) point heat source
\begin{align}
\label{eq:leftHeatSource}
q_{w,1}=W_0\left(1+\frac{1}{\tilde{r}_c}\right)\,\rho_S\, h_m^*\\
\label{eq:rightHeatSource}
q_{w,2}=W_0\left(1-\frac{1}{\tilde{r}_c}\right)\,\rho_S\, h_m^*
\end{align}
The curve radius as a function of the heat flux of the left and right side is then found by substituting \eqref{eq:leftHeatSource} into \eqref{eq:rightHeatSource}, which yields
\begin{align}
\label{eq:analyticalSolutionCurvilinearCCM}
	\tilde{r}_c=\frac{q_{w,1}+q_{w,2}}{q_{w,1}-q_{w,2}}
\end{align}
Equation \eqref{eq:analyticalSolutionCurvilinearCCM} will now serve as the optimal melting reference.

\begin{figure}
	\begin{subfigure}{0.52\textwidth}
		\includegraphics[width=\linewidth]{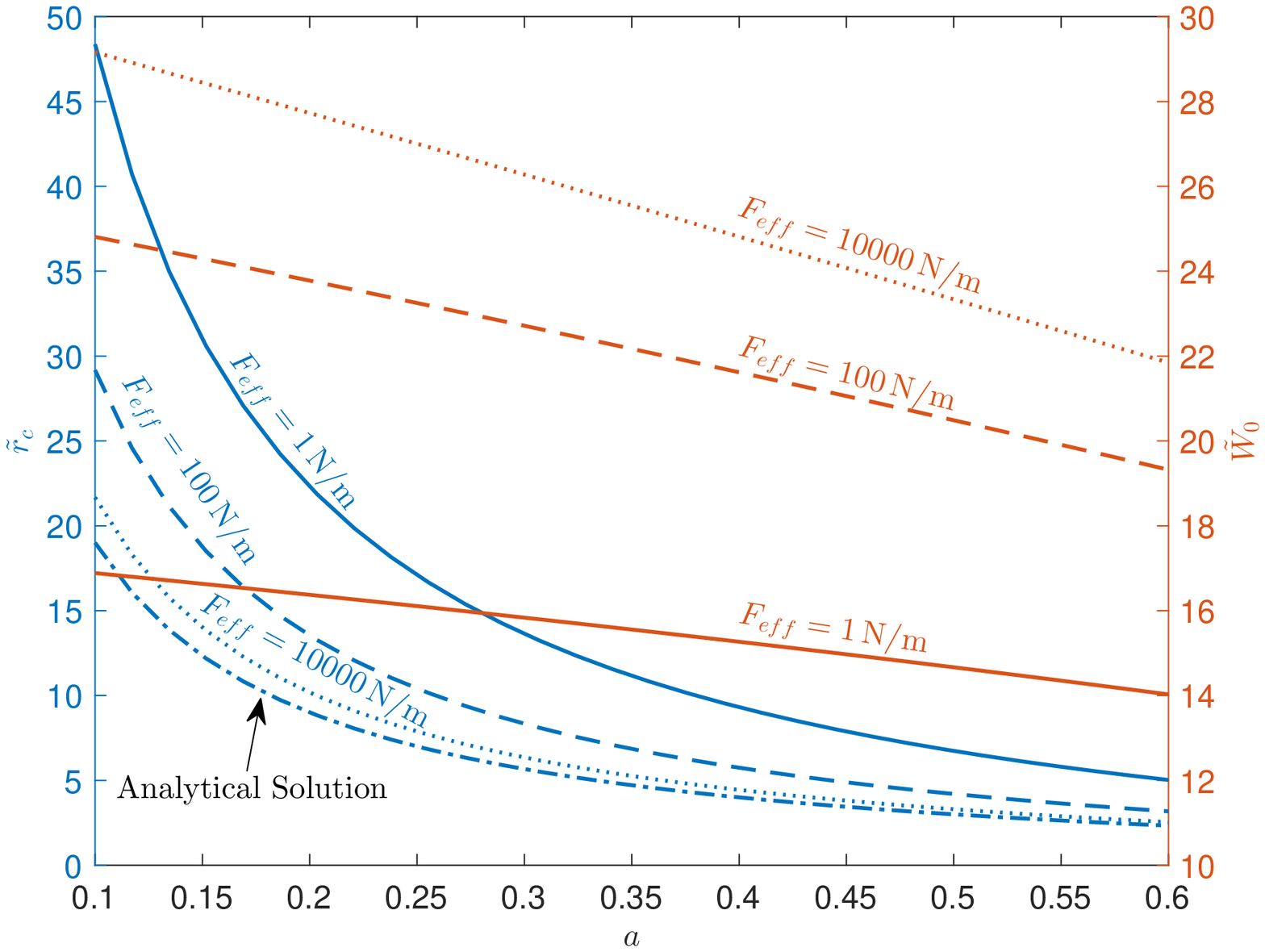}
		\caption{}
		\label{CurvilinearCCMFiguresa}
	\end{subfigure}
	\hspace{-1em}
	\begin{subfigure}{0.52\textwidth}
		\includegraphics[width=\linewidth]{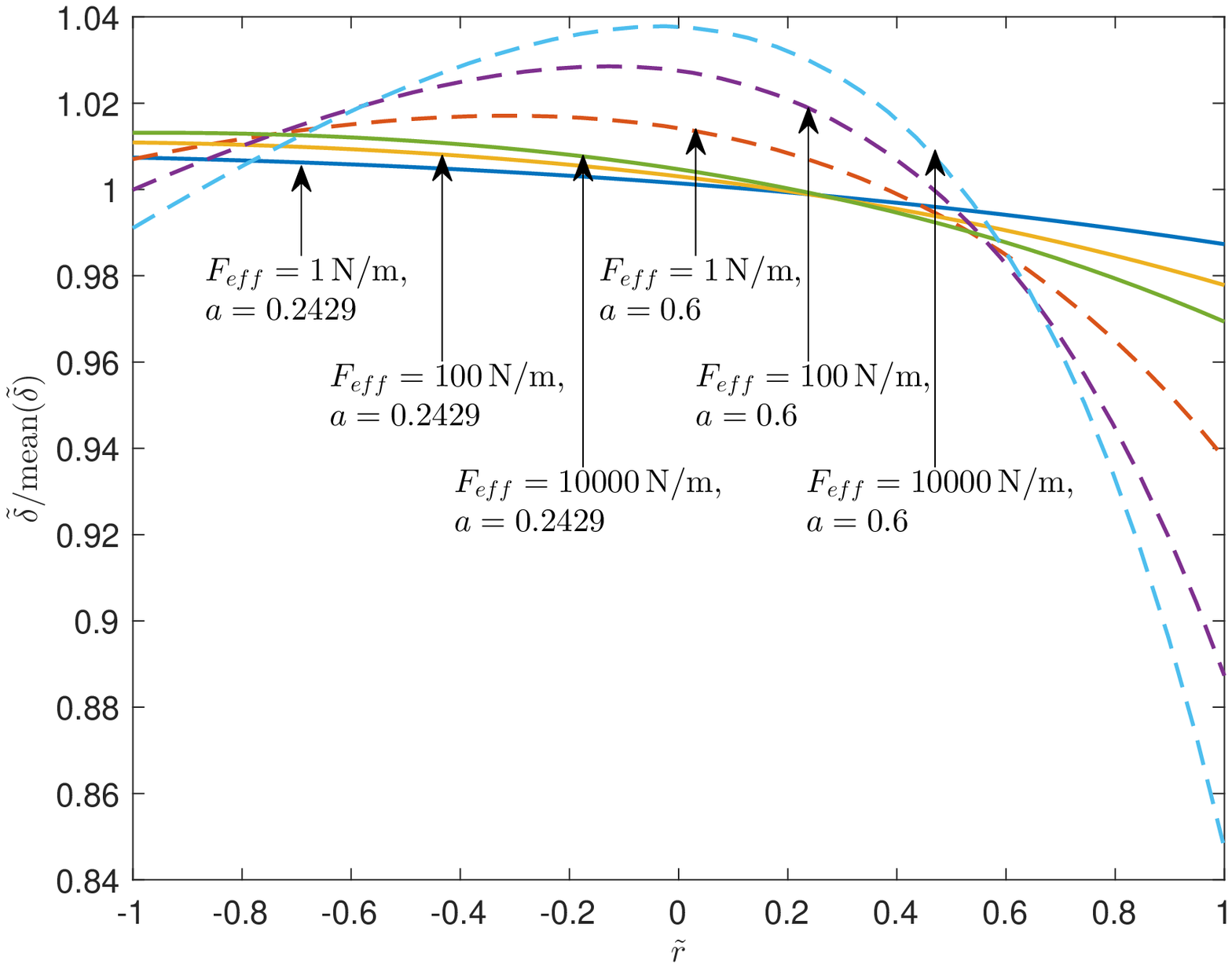}
		\caption{}
		\label{CurvilinearCCMFiguresb}
	\end{subfigure}
	\caption{Curvilinear close-contact melting for a linear heat flux profile that is given by $\tilde{q}''_w=(1-a\tilde{r})/(1-a/2)$ for three different exerted forces $F_\text{eff}$.} \label{CurvilinearCCMFigures}
\end{figure}

Figure \ref{CurvilinearCCMFiguresa} shows the dimensionless curve radii over the heat flux slope factor $a$ for three different contact forces.
The heat flux slope factor $a$ is defined as in the previous section and denotes the slope of the linear heat flux profile according to \eqref{eq:linearHeatFluxProfiles}. In all cases, the maximum heat flux is equal in its value and location ($r=-1$). The heat source will rotate counter-clockwise. For small values of $a$, the curve radius becomes large. As the slope factor vanishes ($a\rightarrow0$), the curve radius tends to infinity $\tilde{r}_c\rightarrow \infty$. This is plausible, as degenerate curve melting with a large curve radius is just straight melting. On the other hand, when $a$ becomes large, the dimensionless radius $\tilde{r}_c$ tends to a finite value $\geq1$ (or $\leq-1$ for clockwise curvilinear close-contact melting). As a consequence to taking convective losses into account, all close-contact curve radii are larger than those obtained with the reference solution \eqref{eq:analyticalSolutionCurvilinearCCM}. For contact forces of $F_\text{eff}=10000\,\text{N}/\text{m}$ and higher, the difference between the computational model and the optimal reference solution becomes very small. The reason for this is that the average melt film thickness decreases with increasing exerted force $F_\text{eff}$, and hence convective losses are less likely to be observed.

The normalized melt film thickness $\tilde{\delta}/\text{mean}(\tilde{\delta})$ for different slope factors $a=0.2429$ and $a=0.6$ is shown in figure \ref{CurvilinearCCMFiguresb}. For $a$ being small, the melt film thickness decreases linearly. For a larger slope factor, the melt film profile has a global maximum that is located close to the heat source center.
This behavior is not intuitive and cannot be determined with the idealized optimal melting relation \eqref{eq:analyticalSolutionCurvilinearCCM}.
	
	\section{Conclusions and Outlook}
In this work we described a numerical method to model quasi-steady close-contact melting, a phenomenon that occurs when a heat source melts its way into a PCM. Our major goal is to determine the relative velocity between the heat source and the PCM, which we call the melting velocity. The proposed method is capable of computing the melting velocity, as well as pressure, velocity and temperature field within the melt film, and the film thickness along the melt film, in a physically consistent way.

In contrast to existing literature that mainly addresses temperature driven close-contact melting or rather simplified heat flux driven situations, our approach is more flexible and allows for a spatially varying heat flux distribution at the working surface. As such, it enables us to study the system's response to varying contact forces in combination with different heat flux distributions. This has not been possible before and constitutes the innovative aspect of our work.

We derived the governing equations based on the balance laws for mass, momentum and energy and reduced its complexity by means of scaling arguments. We furthermore took a Lagrangian perspective, in which the reference frame is fixed to the heat source. The resulting system consists of the steady energy equation coupled to the Reynolds equation via the velocity field. The energy equation is furthermore constraint by local energy balance at the solid-liquid interface, the Stefan condition. Our numerical approach relies on solving both the Reynolds equation, and the energy equation by means of a finite difference discretization. The two solvers are then augmented into an iterative procedure that locally updates the melt film thickness to eventually match the Stefan condition.

We conducted convergence tests and demonstrated proficiency of our procedure by applying it to two different geometries, namely an \caseA\,and a \caseB. For the latter case we extended our approach to rotational melting modes that can be induced by an asymmetric heat flux distribution along the working surface. As part of our results section, we compared results of the computational close-contact melting model with an idealized 'optimal melting' situation that neglects any convective losses. This allowed us to analyze the efficiency of close-contact melting and to quantify the model error introduced if convective losses are neglected.\\ 
Our specific findings are
\begin{enumerate}
	\item General heat flux driven close-contact melting can be computed by means of the straight forward iterative routine sketched in section \ref{sec:solutionalg}.
	\item For a given setting, the melting velocity (or alternatively, the process efficiency) can be increased by concentrating the heat around the center of the heat source.
	\item The curve radius of rotational melting can be decreased by increasing the contact force.
	\item In the limit $F_{eff}\rightarrow\infty$, both melting velocity, and curve radius tend to the optimal melting solution given by \eqref{eq:simpleEnergyBalanceEquation} and \eqref{eq:analyticalSolutionCurvilinearCCM}.
	\item For low Stefan numbers and high forces, the efficiency can be approximated by equation \eqref{eq:approximationVelocityDifference}, in which $P_2\approx-1/3$ and $P_3\approx4/3$ are constant across relevant regimes.
\end{enumerate}
The presented work provides a first step towards a comprehensive computational model for the dynamic behavior of a heat source that melts its way through a material that is undergoing phase-change. Though our approach does consider losses due to convective transport of heat, it is still based on substantial simplifications whose effects need to be studied further. One example is our use of the reduced latent heat of melting instead of solving for the heat equation in the solid domain directly. This is admissible, if the solid temperature initially is close to the melting temperature. But, it is likely to introduce model errors, if the two temperature levels differ significantly. As a next step, we therefore plan to integrate a macro-scale model solving for the heat equation in the solid domain with the hereby presented micro-scale model for the close-contact melt film.

Another future plan is to improve the iterative scheme itself. Within the scope of this paper we utilized a straight forward, local update of the melt film thickness. It turned out to be stable for all our purposes as long as the relaxation factor was chosen appropriately small. However, a less costly and more robust scheme has to be realized, when extending results to more complex geometries, e.g. non-planer, non-axisymmetric, two dimensional working surfaces. Then, it will be necessary to formulate a global scheme, e.g. a global Newton iteration. Moreover, in the context of our application-specific research projects, an important next step will be to conduct a sensitivity analysis of the fitted parameters with respect to geometric setting and thermo-physical regime of interest. Finally, we plan to validate the model with experimental data.

	\section{Acknowledgement}
	The project is supported by the Federal Ministry for Economic Affairs and Energy, Germany, on the basis of a decision by the German Bundestag (FKZ: 50 NA 1206 and 50 NA 1502). It is part of the Enceladus Explorer initiative of the DLR Space Administration.
	
	\section*{References}
	\footnotesize
	\bibliographystyle{elsarticle-harv}
	\bibliography{source}
\end{document}